\documentclass[11pt]{article}
 \usepackage{amsmath,amsthm,amssymb,amsfonts}
\usepackage{mathtools,thmtools}
\usepackage{fullpage}
\usepackage{xcolor}
\usepackage{enumitem}
\usepackage[numbers]{natbib}
\usepackage{todonotes}
\usepackage{bbm}
\usepackage[breaklinks=true,linktocpage=true,hypertexnames=false]{hyperref}
\hypersetup{hidelinks}
\usepackage[nameinlink,noabbrev,capitalize]{cleveref}
\usepackage{float}
\usepackage{tikz}
\usetikzlibrary{positioning,decorations.pathreplacing}
\theoremstyle{plain}
\newtheorem{thm}{Theorem}[section]
\newtheorem{theorem}[thm]{Theorem}
\newtheorem{lemma}[thm]{Lemma}
\newtheorem{fact}[thm]{Fact}

\theoremstyle{definition}

\theoremstyle{remark}
\newtheorem{remark}[thm]{Remark}

\crefname{fact}{Fact}{Facts}
\Crefname{fact}{Fact}{Facts}

\newcommand{\set}[1]{\{#1\}}

\newcommand{\cG}{\mathcal{G}}
\newcommand{\cH}{\mathcal{H}}

\newcommand{\cX}{\mathcal{X}}

\newcommand{\Ldim}{\operatorname{Ldim}}
\newcommand{\tdim}{\operatorname{tdim}}
\newcommand{\VC}{\operatorname{VC}}

\title{Littlestone Dimension and Its Dual}
\author{
  Pooya Hatami\\
  \small Ohio State University\\
  \small \texttt{hatami.2@osu.edu}
  \and
  Chavdar Lalov\\
  \small Ohio State University\\
  \small \texttt{lalov.1@osu.edu}
}

\date{}

\begin{document}
\maketitle

\begin{abstract}
The dual Littlestone dimension of a concept class $\cH\subseteq\{0,1\}^{\cX}$, $\Ldim^*(\cH)\coloneqq\Ldim(\cH^*)$, is the Littlestone dimension of the dual class $\cH^*$, obtained by exchanging the roles of domain points and hypotheses. A classical result of Shelah implies that $\Ldim(\cH)$ and $\Ldim^*(\cH)$ are finite simultaneously, and quantitative results of Hodges~[Mathematika, 1981] imply the doubly exponential bound $\Ldim(\cH)\le 2^{2^{O(\Ldim^*(\cH))}}$. Malliaris and Moran~[Trans.\ Amer.\ Math.\ Soc., 2024] asked whether this bound is optimal. We answer this by proving the bound
\[
\Ldim(\cH)\le 2^{O(\Ldim^*(\cH))},
\]
which is tight up to the constant in the exponent, as witnessed by the class of all binary hypotheses on a finite domain. 
\end{abstract}

\section{Introduction}
The \emph{Littlestone dimension} $\Ldim(\cH)$ of a class
$\cH\subseteq\{0,1\}^{\cX}$ is the largest depth of a
complete binary tree, with internal nodes labeled by points
of $\cX$, all of whose root-to-leaf paths are realized by
hypotheses in $\cH$.
Remarkably, the finiteness of Littlestone dimension
characterizes fundamental dichotomies in two seemingly
unrelated fields: model theory and learning theory.
In model theory, Littlestone dimension is an instance of Shelah's $2$-rank~\cite{Shelah90}, and a formula is stable precisely when its associated concept class has finite Littlestone dimension.
In learning theory, Littlestone~\cite{Littlestone88}
introduced this dimension and showed that it equals the
optimal worst-case mistake bound for deterministic online
learning in the realizable setting.
Its finiteness also characterizes agnostic online
learning~\cite{BPS09}, approximate differentially-private PAC learning~\cite{ALMM19,BLM20}, and
list-replicable learning~\cite{CMY23}.

The connection between the model-theoretic and
learning-theoretic viewpoints has been explored explicitly
in recent work~\cite{ChaseFreitag19,Bhaskar21,MalliarisMoranUFT,malliaris2025epsilon},
revealing fruitful interactions between the two fields.
In particular, stability theory provides a rich source of
classes of finite Littlestone dimension, while the
combinatorial and algorithmic properties of such classes play a central
role in regularity lemmas for stable
graphs~\cite{MalliarisShelah14,malliaris2024model,coregliano2026stable}.

Given a class $\cH\subseteq \{0,1\}^\cX$, one can exchange the roles of domain points and hypotheses. As a result, every point $x\in \cX$ determines a function on $\cH$, mapping $h\mapsto h(x)$. The class of all these functions is the dual class $\cH^*\subseteq \{0,1\}^\cH$, and $(\cH^*)^*$ is the same object as $\cH$, up to natural identification.   
In learning theory, the complexity of the dual class matters when hypotheses are sampled rather than points, as in sample compression schemes, whose size is controlled by the VC dimensions of $\cH$ and $\cH^*$ together~\cite{MoranYehudayoff16}. It is natural to ask how the complexity of $\cH^*$ compares with that of $\cH$. For VC dimension, this is settled: Assouad~\cite{Assouad83} showed that $\VC(\cH^*)\le 2^{\VC(\cH)+1}-1$, and this is tight.

For Littlestone dimension, write $\Ldim^*(\cH)\coloneqq\Ldim(\cH^*)$. Shelah's unstable formula theorem~\cite{Shelah90} implies that $\Ldim(\cH)$ and $\Ldim^*(\cH)$ are finite together. The proof goes through a third quantity, the \emph{threshold dimension} $\tdim(\cH)$, that is invariant under duality. It is the length of the longest sequence of points $x_1,\dots,x_k$ and hypotheses $h_1,\dots,h_k$ with $h_i(x_j)=1$ if and only if $i\le j$. Hodges~\cite{Hodges81} made the relationship between threshold dimension and Littlestone dimension quantitative (see also~\cite[Theorem~3]{ALMM19}). For every class $\cH$ of positive finite Littlestone dimension,
\begin{equation}\label{eq:hodges}
\lfloor\log_2\Ldim(\cH)\rfloor\ \le\ \tdim(\cH)\ \leq\ 2^{\Ldim(\cH)+1}-1.
\end{equation}
As noted in
\cite[Corollary~3.6]{Bhaskar21}, combining these
bounds through the dual gives
\[
\Ldim(\cH)\le 2^{2^{O(\Ldim^*(\cH))}}.
\]
Whether this double exponential bound is tight was asked in \cite[Section 7]{malliaris2024model} (see also \cite{Bhaskar21}). We answer this question in the negative by proving a single exponential bound that is tight up to a constant factor in the exponent.

\begin{theorem}\label{thm:main}
For every concept class $\cH\subseteq\{0,1\}^{\cX}$, we have
\[
\Ldim(\cH)\le\frac{3}{2}\cdot 9^{\Ldim^*(\cH)}=  2^{O(\Ldim^*(\cH))}.
\]
\end{theorem}
The class of all binary hypotheses on $2^d$ points has 
Littlestone dimension $2^d$ and dual Littlestone dimension $d$, so \cref{thm:main} is tight up to the base of the exponential.

The proof of \cref{thm:main} builds upon a classical Ramsey lemma
for binary trees (\cref{fact:ramsey}) to establish a new structural result for classes of large
Littlestone dimension. Each path in a shattered tree partitions
its points into two fibers, according to whether it takes the
$0$- or $1$-edge. We prove a path lemma (\cref{lem:path})
guaranteeing a path for which the sum of the dual Littlestone
dimensions of these fibers is at least logarithmic in the
tree depth. This lower bound on the sum, together with the
standard recursion property of Littlestone dimension
(\cref{fact:recursion}), forces $\cH$ to have large dual
Littlestone dimension and yields the exponential bound.

\paragraph{Version 1.}In the first version of this paper, we proved the weaker bound $\Ldim(\cH)\le 2^{O(\Ldim^*(\cH)^2)}$. After it appeared, Yuval Filmus shared with us a proof of \cref{thm:main} that had been generated by an AI model. We present that proof in \cref{sec:exp}. We present our original argument in \cref{sec:dich}, since it establishes a structural dichotomy for classes of large Littlestone dimension that may be of independent interest. The two proofs share their basic ingredients, a Ramsey lemma for two-colored binary trees and a recursion for the dual Littlestone dimension \eqref{eq:dualrec}.

Our original proof uses subadditivity of Littlestone dimension under domain partitions to establish a combinatorial
dichotomy for classes of large Littlestone dimension (\cref{lem:dich}).
Suppose $\Ldim(\cH)\ge k\ell$ for positive integers $k,\ell$.
Either the class contains a hypothesis whose fibers partition the domain
into two sets, each of which retains Littlestone dimension at least $\ell$,
or the class contains a threshold sequence of length $k$.
In the first case, we use induction to obtain a
lower bound on the dual Littlestone dimension,
while in the second case such a bound follows
directly from the threshold sequence.
Combining these two alternatives yields the bound $2^{O(\Ldim^*(\cH)^2)}$.

\paragraph{AI methodology.}
The authors developed the main proof strategy of \cref{sec:dich}, combining induction on the dual Littlestone dimension with the construction of threshold configurations. They used AI assistants to produce a first version of the proof and then refined the argument through several rounds of discussion with the assistants. The proof in \cref{sec:exp} was generated by ChatGPT, prompted by Yuval Filmus with the first version of this paper. The authors verified it and rewrote it in the notation of this paper. AI assistance was also used in editing the exposition and locating references. The authors reviewed all AI-assisted revisions, edited them as needed, and made all final decisions about the proof and its presentation.

\paragraph{Acknowledgment.}
The authors thank Ari Blondal, Hamed Hatami, and Sivan Tretiak for helpful discussions. We thank Yuval Filmus for sharing the proof presented in \cref{sec:exp} and allowing us to include it in this note. 

\section{Preliminaries}

Throughout, $\cX$ is a (possibly infinite) set and
$\cH\subseteq\{0,1\}^{\cX}$ is a \emph{binary concept class}.
The elements of $\cH$ are called \emph{concepts}.
It is often convenient to think of $\cH$ as a binary matrix with rows
labeled by $\cH$ and columns labeled by $\cX$ (\cref{fig:shattered-tree}).
For $\cX'\subseteq\cX$, write
\[
\cH|_{\cX'}\coloneqq\{h|_{\cX'}:h\in\cH\}.
\]
For $\cG\subseteq\cH$ and $\cX'\subseteq\cX$, we call
$\cG|_{\cX'}$ a \emph{submatrix} of $\cH$.

For $x\in\cX$, $h\in\cH$, and $b\in\{0,1\}$, write
\[
\cH^{x,b}\coloneqq\{h\in\cH:h(x)=b\},
\qquad
h^{-1}(b)\coloneqq\{x\in\cX:h(x)=b\}.
\]

The \emph{dual class} of $\cH$ is
$\cH^*\coloneqq\{x^*:x\in\cX\}\subseteq\{0,1\}^{\cH}$,
where $x^*(h)=h(x)$.
Viewing $\cH$ as a matrix, duality corresponds to
simply transposing the matrix.

\paragraph{Littlestone dimension.}
We consider complete binary trees whose internal nodes are
labeled by elements of $\cX$, with the two outgoing edges at
each internal node labeled $0$ and $1$.
Such a tree is \emph{shattered} by $\cH$ if every root-to-leaf
path is realized by some $h\in\cH$: whenever the path takes
an edge labeled $b$ from a node labeled $x$, we have $h(x)=b$ (see \cref{fig:shattered-tree}).
For $\cH\ne\emptyset$, the \emph{Littlestone dimension} of
$\cH$~\cite{Littlestone88}, denoted $\Ldim(\cH)$, is the maximum depth
of a finite complete binary tree shattered by $\cH$, or $\infty$ if no
maximum exists.
We set $\Ldim(\emptyset)=-1$, and define the Littlestone dimension of any
singleton class to be zero.

The \emph{dual Littlestone dimension} of $\cH$ is
$\Ldim^*(\cH)\coloneqq\Ldim(\cH^*)$.

Finally, we remark that Littlestone dimension is also known as
thicket dimension~\cite{Bhaskar21} and is an
instance of Shelah's $2$-rank in model
theory~\cite{Shelah90,ChaseFreitag19}.

\begin{figure}[H]
\centering

\begin{minipage}[t]{0.53\textwidth}
\vspace{0pt}
\centering
\begin{tikzpicture}[
    x=0.95cm,
    y=0.95cm,
    point/.style={draw,circle,fill=white,
        inner sep=1.5pt,font=\scriptsize},
    leaf/.style={draw,circle,fill=black,inner sep=1.2pt},
    lab/.style={font=\scriptsize,fill=white,
        inner sep=0.8pt}
]
    \node[point] (v) at (0,0) {$x_1$};

    \node[point] (v0) at (-2,-1) {$x_2$};
    \node[point] (v1) at (2,-1) {$x_3$};

    \node[point] (v00) at (-3,-2) {$x_4$};
    \node[point] (v01) at (-1,-2) {$x_5$};
    \node[point] (v10) at (1,-2) {$x_6$};
    \node[point] (v11) at (3,-2) {$x_7$};

    \node[leaf,label=below:{\scriptsize $h_{000}$}]
        (l000) at (-3.5,-3) {};
    \node[leaf,label=below:{\scriptsize $h_{001}$}]
        (l001) at (-2.5,-3) {};
    \node[leaf,label=below:{\scriptsize $h_{010}$}]
        (l010) at (-1.5,-3) {};
    \node[leaf,label=below:{\scriptsize $h_{011}$}]
        (l011) at (-0.5,-3) {};
    \node[leaf,label=below:{\scriptsize $h_{100}$}]
        (l100) at (0.5,-3) {};
    \node[leaf,label=below:{\scriptsize $h_{101}$}]
        (l101) at (1.5,-3) {};
    \node[leaf,label=below:{\scriptsize $h_{110}$}]
        (l110) at (2.5,-3) {};
    \node[leaf,label=below:{\scriptsize $h_{111}$}]
        (l111) at (3.5,-3) {};

    \draw (v) -- node[lab,midway,above left=2pt] {$0$} (v0);
    \draw (v) -- node[lab,midway,above right=2pt] {$1$} (v1);

    \draw (v0) -- node[lab,midway,above left=2pt] {$0$} (v00);
    \draw (v0) -- node[lab,midway,above right=2pt] {$1$} (v01);
    \draw (v1) -- node[lab,midway,above left=2pt] {$0$} (v10);
    \draw (v1) -- node[lab,midway,above right=2pt] {$1$} (v11);

    \draw (v00) -- node[lab,midway,above left=2pt] {$0$} (l000);
    \draw (v00) -- node[lab,midway,above right=2pt] {$1$} (l001);
    \draw (v01) -- node[lab,midway,above left=2pt] {$0$} (l010);
    \draw (v01) -- node[lab,midway,above right=2pt] {$1$} (l011);
    \draw (v10) -- node[lab,midway,above left=2pt] {$0$} (l100);
    \draw (v10) -- node[lab,midway,above right=2pt] {$1$} (l101);
    \draw (v11) -- node[lab,midway,above left=2pt] {$0$} (l110);
    \draw (v11) -- node[lab,midway,above right=2pt] {$1$} (l111);
\end{tikzpicture}
\end{minipage}%
\hfill
\begin{minipage}[t]{0.44\textwidth}
\vspace{0pt}
\centering
\footnotesize
\setlength{\tabcolsep}{3pt}
\renewcommand{\arraystretch}{1.25}
\begin{tabular}{c|ccccccc}
 & $x_1$ & $x_2$ & $x_3$ & $x_4$ & $x_5$ & $x_6$ & $x_7$\\
\hline
$h_{000}$ & 0 & 0 & 0 & 0 & 0 & 0 & 0\\
$h_{001}$ & 0 & 0 & 0 & 1 & 0 & 0 & 0\\
$h_{010}$ & 0 & 1 & 0 & 0 & 0 & 0 & 0\\
$h_{011}$ & 0 & 1 & 0 & 0 & 1 & 0 & 0\\
$h_{100}$ & 1 & 0 & 0 & 0 & 0 & 0 & 0\\
$h_{101}$ & 1 & 0 & 0 & 0 & 0 & 1 & 0\\
$h_{110}$ & 1 & 0 & 1 & 0 & 0 & 0 & 0\\
$h_{111}$ & 1 & 0 & 1 & 0 & 0 & 0 & 1
\end{tabular}
\end{minipage}

\caption{A complete binary tree of depth $3$ shattered by
a concept class, together with its evaluation matrix.
Rows correspond to hypotheses and columns to domain points.
Each hypothesis $h_{abc}$ realizes the path with edge
labels $a,b,c$. Entries outside the corresponding path are chosen arbitrarily (here they are set
to $0$ for simplicity).}
\label{fig:shattered-tree}
\end{figure}

\paragraph{Threshold dimension.}
A \emph{threshold sequence} of length $k$ consists of
hypotheses $h_1,\ldots,h_k\in\cH$ and points
$x_1,\ldots,x_k\in\cX$ such that
\[
h_i(x_j)=1
\quad\Longleftrightarrow\quad
i\le j.
\]
The \emph{threshold dimension} of $\cH$~\cite{ALMM19},
denoted $\tdim(\cH)$, is the maximum $k\ge0$ for which
$\cH$ contains a threshold sequence of length $k$, or $\infty$ if no maximum exists.
Note that
\[
\tdim(\cH)=\tdim(\cH^*).
\]

Threshold dimension is a finite quantitative counterpart
of the order property from model theory~\cite{Shelah90}.
Related configurations appear as ladders in
Hodges~\cite{Hodges81} and as half-graphs in
Malliaris--Shelah~\cite{MalliarisShelah14}.

\subsection{Basic properties of Littlestone dimension}
We collect a few basic but useful properties of the Littlestone dimension. First is the monotonicity of the Littlestone dimension under restrictions. For every $\cG\subseteq\cH$ and $\cX'\subseteq\cX$, we have
\[
\Ldim(\cG|_{\cX'})\le\Ldim(\cH).
\]
Hence, we also obtain monotonicity for the dual Littlestone dimension.

The following is the standard recursion for Littlestone dimension that is used in Littlestone's optimal online learning algorithm~\cite{Littlestone88}.
\begin{fact}[Littlestone recursion]\label{fact:recursion}
Let $\cH \subseteq \set{0,1}^{\cX}$ be a nonempty concept class
with $d\coloneqq\Ldim(\cH)<\infty$.
For every $x\in\cX$,
\[
\min_{b\in\{0,1\}}\Ldim(\cH^{x,b})\le d-1.
\]
Moreover, if $\cX$ is nonempty, equality holds for some
$x\in\cX$.
\end{fact}

\begin{proof}
If both subclasses $\cH^{x,0}$ and $\cH^{x,1}$ had
Littlestone dimension at least $d$, we could combine
their shattered trees beneath a root labeled $x$,
obtaining a shattered tree of depth $d+1$.
This proves the first inequality.

For $d\ge1$, taking $x$ to be the root of a shattered
tree of depth $d$ gives the reverse inequality.
If $d=0$, one of the two
subclasses is empty for every $x\in\cX$, so equality
also holds.
\end{proof}

It is useful to apply this recursion to the dual class. Let $\cX\ne\emptyset$ and $h\in\cH$. The hypotheses of $(\cH^*)^{h,b}$ are the evaluation maps $x^*$ with $x\in h^{-1}(b)$, so $(\cH^*)^{h,b}$ is the dual class of $\cH|_{h^{-1}(b)}$, up to duplicate domain points. \cref{fact:recursion}, applied to $\cH^*$ at the point $h$ when $\Ldim^*(\cH)<\infty$, gives
\begin{equation}\label{eq:dualrec}
\Ldim^*(\cH)\ \ge\ 1+\min_{b\in\{0,1\}}\Ldim^*\bigl(\cH|_{h^{-1}(b)}\bigr),
\end{equation}
and this holds trivially when $\Ldim^*(\cH)=\infty$.

In \cref{sec:exp}, we sometimes pass to smaller trees that are \emph{embedded} inside a complete binary tree. We say that a complete binary tree $T'$ is embedded in a complete binary tree $T$ if the nodes of $T'$ are a subset of the nodes of $T$, the leaves of $T'$ are leaves of $T$, and for every internal node $v$ of $T'$ and every $b\in\{0,1\}$, the child of $v$ in direction $b$ in $T'$ lies in the subtree of $T$ below the edge labeled $b$ at $v$. Note that every root-to-leaf path of $T'$ extends to some root-to-leaf path of $T$ that takes the same edges at the nodes of $T'$. Thus, if $T$ is shattered by $\cH$, then so is $T'$. 

We use the following Ramsey-type lemma for two-colored binary trees.
See \cite[Lemma~16]{ALMM19} for a short inductive proof.

\begin{fact}[Tree Ramsey lemma]\label{fact:ramsey}
Let $p,q\ge1$ be integers. If the internal nodes of a complete binary tree of depth $p+q-1$ are colored red and blue, then the tree contains an embedded tree of depth $p$ all of whose internal nodes are red, or an embedded tree of depth $q$ all of whose internal nodes are blue.
\end{fact}

\cref{sec:dich} uses the following subadditivity property, which is a consequence of \cref{fact:ramsey}. We include a self-contained direct proof.

\begin{fact}[Subadditivity]\label{fact:subadd}
For every nonempty $\cH\subseteq \{0,1\}^\cX$ and every $\cX'\subseteq \cX$,
\[
\Ldim(\cH)\le \Ldim(\cH|_{\cX'})+ \Ldim(\cH|_{\cX\backslash \cX'}). 
\]
\end{fact}
\begin{proof}

If either term on the right is infinite, there is nothing to prove, so let $d\coloneqq\Ldim(\cH|_{\cX'})+\Ldim(\cH|_{\cX\setminus\cX'})<\infty$. We show $\Ldim(\cH)\le d$ by induction on $d$.

The base case $d=0$ is trivial, since in this case both restrictions are singleton classes and thus $\Ldim(\cH)=0$. Let $d\ge1$ and assume towards contradiction that $\Ldim(\cH)\ge d+1$. Let $x$ be the label of the root of a tree of depth $d+1$ shattered by $\cH$, so that $\Ldim(\cH^{x,b})\ge d$ for both $b\in\{0,1\}$. By symmetry, we may assume $x\in\cX'$. By \cref{fact:recursion} applied to $\cH|_{\cX'}$, there is $b\in\{0,1\}$ with $\Ldim\bigl(\cH^{x,b}|_{\cX'}\bigr)=\Ldim\bigl((\cH|_{\cX'})^{x,b}\bigr)\le\Ldim(\cH|_{\cX'})-1$. Together with monotonicity under restriction, we have \[ \Ldim(\cH^{x,b}|_{\cX'})+\Ldim(\cH^{x,b}|_{\cX\setminus\cX'})\le\bigl(\Ldim(\cH|_{\cX'})-1\bigr)+\Ldim(\cH|_{\cX\setminus\cX'})=d-1, \] so the induction hypothesis gives $\Ldim(\cH^{x,b})\le d-1$, contradicting $\Ldim(\cH^{x,b})\ge d$. \end{proof}

Finally, we include a short proof of the second inequality in \cref{eq:hodges} that we will use.

\begin{fact}[Thresholds and Littlestone dimension]\label{fact:thr}
If $\cH$ contains a threshold sequence of length $k\ge1$,
then
\[
\Ldim(\cH)\ge\lfloor\log_2 k\rfloor.
\]
\end{fact}
\begin{proof}
Let $h_1,\dots, h_k$ and $x_1,\dots,x_k$ be a threshold sequence of $\cH$.
Querying $x_m$ splits $\{h_1,\dots,h_k\}$ into $\{h_i: i\le m\}$ and $\{h_i: i>m\}$, which are threshold sequences on $\{x_j: j\le m\}$ and $\{x_j: j>m\}$. Setting $m=\lfloor k/2\rfloor$ and iterating, this gives a shattered tree of depth $\lfloor\log_2k\rfloor$.
\end{proof}

Since $\tdim(\cH)=\tdim(\cH^*)$, \cref{fact:thr} applied to $\cH^*$ shows that a threshold sequence of length $k\ge1$ in $\cH$ also gives $\Ldim^*(\cH)\ge\lfloor\log_2k\rfloor$.

\section{Proof of \texorpdfstring{\cref{thm:main}}{the main theorem}}\label{sec:exp}
Let $d=\Ldim^*(\cH)$. We prove that if $0\le d<\infty$, then $\Ldim(\cH)\le \frac{3^{2d+1}-1}{2}.$

For a root-to-leaf path $P$ of a tree with internal nodes labeled by points of $\cX$ and $b\in\{0,1\}$, let $X_b(P)$ denote the set of labels of the nodes at which $P$ takes the edge labeled $b$. If the tree is shattered by $\cH$, then any hypothesis realizing $P$ is identically $b$ on $X_b(P)$.

For $j\geq0$ define 
\[N_j=\frac{3^{j+1}+1}{2},\]
and note $N_{j+1}=3N_j-1$.
The heart of the proof is the following lemma.
\begin{lemma}[Path lemma]\label{lem:path}
Let $j\ge0$. Every complete binary tree $T$ of depth $N_j$ shattered by $\cH$ has a root-to-leaf path $P$ such that $X_0(P)$ and $X_1(P)$ are nonempty and
\[
\Ldim^*\bigl(\cH|_{X_0(P)}\bigr)+\Ldim^*\bigl(\cH|_{X_1(P)}\bigr)\ \ge\ j.
\]
\end{lemma}
\begin{proof}
We proceed by induction on $j$. For the base case $j=0$, we can simply choose the path
$P$ that follows the $0$-edge at the root and the $1$-edge at the next level. 

For the induction step, suppose $j\ge 0$, and let $T$ be a tree of depth $N_{j+1}=3N_j-1$ shattered by $\cH$. The top $N_j$ levels of $T$ form a complete binary tree of depth $N_j$. By the induction hypothesis, it has a root-to-leaf path $R$ with $R_b\coloneqq X_b(R)$ nonempty for both $b$ and \begin{equation}\label{eq:pair_1}\Ldim^*(\cH|_{R_0})+\Ldim^*(\cH|_{R_1})\ge j.\end{equation}Fix a hypothesis $h\in\cH$ realizing $R$, so that $h$ equals $b$ on $R_b$.

Let $T_1$ be the subtree of $T$ below the endpoint of $R$ and note that it has depth $2N_j-1$. Color each internal node of $T_1$ labeled $x$ according to the value of $h(x)$. By \cref{fact:ramsey}, $T_1$ contains an embedded tree $T_2$ of depth $N_j$ all of whose node labels lie in $h^{-1}(c)$ for some $c\in\{0,1\}$. Since $T_2$ is shattered by $\cH$, the induction hypothesis implies that it has a root-to-leaf path $Q$ with $Q_b\coloneqq X_b(Q)$ nonempty for both $b$ and \begin{equation}\label{eq:pair_2}    
\Ldim^*(\cH|_{Q_0})+\Ldim^*(\cH|_{Q_1})\ge j.\end{equation}
Note that $Q_0,Q_1\subseteq h^{-1}(c)$.

Extend $Q$ to a root-to-leaf path of $T_1$ that takes
edges with the same labels at the internal nodes of
$T_2$ along $Q$. Let $P$ be
the root-to-leaf path of $T$ obtained by following $R$
and then this extension. Notice $X_b(P)\supseteq R_b\cup Q_b$ for both $b$. Put
\[
m_b\coloneqq\max\bigl\{\Ldim^*(\cH|_{R_b}),\ \Ldim^*(\cH|_{Q_b})\bigr\},
\]
so that $\Ldim^*(\cH|_{X_b(P)})\ge m_b$ by monotonicity. If $m_0+m_1\ge j+1$, we are done.

Otherwise $m_0+m_1\le j$. Hence, by \eqref{eq:pair_1} and \eqref{eq:pair_2} it follows that $m_0+m_1=j$ and 
\[
\Ldim^*(\cH|_{R_0})=\Ldim^*(\cH|_{Q_0})=m_0,
\qquad
\Ldim^*(\cH|_{R_1})=\Ldim^*(\cH|_{Q_1})=m_1.
\]
 Put $b\coloneqq1-c$. Then $h$ is identically $b$ on $R_b$ and identically $c=1-b$ on $Q_b$, so the fibers of $h|_{R_b\cup Q_b}$ are $R_b$ and $Q_b$. As a result, the dual recursion \eqref{eq:dualrec} applied to the class $\cH|_{R_b\cup Q_b}$ gives
\[
\Ldim^*\bigl(\cH|_{X_b(P)}\bigr)\ \ge\ \Ldim^*(\cH|_{R_b\cup Q_b})\ \ge\ 1+\min\{\Ldim^*(\cH|_{R_b}),\Ldim^*(\cH|_{Q_b})\}\ =\ m_b+1 .
\]
Together with $\Ldim^*(\cH|_{X_{1-b}(P)})\ge m_{1-b}$, this yields $\Ldim^*(\cH|_{X_0(P)})+\Ldim^*(\cH|_{X_1(P)})\ge j+1$.
\end{proof}

\begin{proof}[Proof of \cref{thm:main}]
Let $d\coloneqq \Ldim^*(\cH)$. When $d=\infty$ there is nothing to prove, and when $d=-1$ we have $\Ldim(\cH)\le0 \le \frac{3}{2}\cdot 9^d=\frac{1}{6}$. 

So let $0\le d<\infty$, and suppose for the sake of contradiction that $\cH$ shatters a complete binary tree $T$ of depth $N_{2d}$. By \cref{lem:path}, $T$ has a root-to-leaf path $P$ with $X_0\coloneqq X_0(P)$ and $X_1\coloneqq X_1(P)$ nonempty and $\Ldim^*(\cH|_{X_0})+\Ldim^*(\cH|_{X_1})\ge2d$. By monotonicity, both $\Ldim^*(\cH|_{X_0})$ and $\Ldim^*(\cH|_{X_1})$ are at most $d$, and thus both equal $d$. A hypothesis $h\in\cH$ realizing $P$ is identically $b$ on $X_b$, so the fibers of $h|_{X_0\cup X_1}$ are $X_0$ and $X_1$. Consequently, \eqref{eq:dualrec} applied to $\cH|_{X_0\cup X_1}$ gives $\Ldim^*(\cH|_{X_0\cup X_1})\ge d+1$, which contradicts $\Ldim^*(\cH|_{X_0\cup X_1})\le\Ldim^*(\cH)=d$. Hence,
\[
\Ldim(\cH)\ \le\ N_{2d}-1\ =\ \frac{3^{2d+1}-1}{2} < \frac{3}{2}\cdot 9^d.\qedhere
\]
\end{proof}

\section{\texorpdfstring{A dichotomy lemma}{A dichotomy lemma}}\label{sec:dich}
In this section we present our original proof of the bound $\Ldim(\cH)\le2^{O(\Ldim^*(\cH)^2)}$.

The heart of the proof is a dichotomy:  a class of large Littlestone dimension either contains a submatrix with a hypothesis both of whose fibers have large Littlestone dimension, or it contains a long \emph{extended threshold}. An extended threshold of size $k\ge0$ in $\cH$ consists of hypotheses $h_0,\dots,h_k\in\cH$ and points $x_1,\dots,x_k\in\cX$ such that $h_i(x_j)=1$ iff $i<j$. Dropping $h_k$ from an extended threshold of size $k$ leaves a threshold sequence of length $k$.

\begin{lemma}[Dichotomy]\label{lem:dich}
Let $\ell\ge1$ and $k\ge0$ be integers, and let $\cH\subseteq\{0,1\}^\cX$ be a class with $\Ldim(\cH)\ge k\ell$. Then at least one of the following holds.
\begin{enumerate}[label=(\alph*)]
\item\label{it:a} Some submatrix $\cG$ of $\cH$ contains a hypothesis $g\in\cG$ with $\Ldim\bigl(\cG|_{g^{-1}(b)}\bigr)\ge \ell$ for both $b\in\{0,1\}$.
\item\label{it:b} $\cH$ contains an extended threshold of size $k$.
\end{enumerate}
\end{lemma}

\begin{proof}
By passing to a finite subclass that shatters a tree of depth
$k\ell$, we may assume $\Ldim(\cH)<\infty$. We proceed by induction on $k$. For $k=0$, \ref{it:b} holds with any $h_0\in\cH$.

Let $k\ge1$ and fix any $h\in\cH$. If $\Ldim\bigl(\cH|_{h^{-1}(b)}\bigr)\ge \ell$ for both $b\in\{0,1\}$, then \ref{it:a} holds with $\cG=\cH$ and $g=h$. Otherwise there is $b\in\{0,1\}$ with
\[
\Ldim\bigl(\cH|_{h^{-1}(1-b)}\bigr)\le \ell-1.
\]
Put $\cX'\coloneqq h^{-1}(b)$. By subadditivity (\cref{fact:subadd}),
\[
\Ldim(\cH|_{\cX'})\ \ge\ k\ell-(\ell-1)\ \ge\ 1,
\]
so by \cref{fact:recursion} applied to $\cH|_{\cX'}$ there is $x_0\in\cX'$ with
\[
\Ldim\bigl((\cH|_{\cX'})^{x_0,1-b}\bigr)\ \ge\ k\ell-\ell\ =\ (k-1)\ell.
\]

\begin{figure}[H]
\centering
\begin{minipage}{13cm}
\centering
\begin{tikzpicture}[
    x=0.93cm,
    y=0.78cm,
    entry/.style={font=\small, inner sep=0pt},
    label/.style={font=\small, inner sep=0pt},
    annot/.style={font=\footnotesize, inner sep=0pt}
]
    \def\L{-0.25}
    \def\R{10.90}
    \def\S{7.00}
    \def\T{5.65}
    \def\B{-1.55}
    \draw[line width=0.6pt] (\L,\B) rectangle (\R,\T);
    \draw[line width=1.0pt] (\S,\B) -- (\S,\T);
    \draw[line width=0.65pt] (\L,2.95) -- (\R,2.95);
    \draw[line width=0.65pt] (\L,3.85) -- (\R,3.85);

    \draw[densely dashed,line width=0.5pt] (0.30,\B) -- (0.30,\T);
    \draw[densely dashed,line width=0.5pt] (1.40,\B) -- (1.40,\T);

    \node[label] at (0.85,6.02) {$x_0$};
    \node[label] at (2.12,6.02) {$z_1$};
    \node[label] at (3.40,6.02) {$z_2$};
    \node[label] at (4.68,6.02) {$z_3$};
    \node[label] at (5.95,6.02) {$\cdots$};
    \node[label] at (8.65,6.02) {$\cdots$};

    \draw[decorate,decoration={brace,amplitude=4pt}] (\L,6.40)--(\S,6.40)
      node[midway,above=5pt,label] {$\mathcal{X}'=h^{-1}(1)$};
    \draw[decorate,decoration={brace,amplitude=4pt}] (\S,6.40)--(\R,6.40)
      node[midway,above=5pt,label] {$\mathcal{X}\setminus \mathcal{X}'$};

    \node[label,left=5pt] at (\L,5.2) {$\vdots$};
    \node[label,left=5pt] at (\L,4.3) {$\vdots$};
    \node[label,left=5pt] at (\L,3.4) {$h$};
    \node[label,left=5pt] at (\L,2.5) {$g_0$};
    \node[label,left=5pt] at (\L,1.6) {$g_1$};
    \node[label,left=5pt] at (\L,0.7) {$g_2$};
    \node[label,left=5pt] at (\L,-0.2) {$g_3$};
    \node[label,left=5pt] at (\L,-1.1) {$\vdots$};

    \foreach \y in {5.2,4.3}{
      \node[entry] at (0.85,\y) {$1$};
      \foreach \x in {2.12,3.40,4.68,5.95,8.65,9.95}
        {\node[entry] at (\x,\y) {$\cdot$};}
    }

    \foreach \x in {0.85,2.12,3.40,4.68,6.55}
      {\node[entry] at (\x,3.4) {$1$};}
    \node[entry] at (5.95,3.4) {$\cdots$};
    \node[entry] at (7.70,3.4) {$0$};
    \node[entry] at (8.65,3.4) {$\cdots$};
    \node[entry] at (10.00,3.4) {$0$};

    \foreach \y in {2.5,1.6,0.7,-0.2}{
      \node[entry] at (0.85,\y) {$0$};
      \node[entry] at (5.95,\y) {$\cdots$};
      \node[entry] at (8.65,\y) {$\cdots$};
    }
    \node[entry] at (0.85,-1.1) {$\vdots$};
    \node[entry] at (3.40,-1.1) {$\cdots$};
    \node[entry] at (8.65,-1.1) {$\cdots$};

    \foreach \y/\a/\b/\c in {2.5/1/1/1,1.6/0/1/1,0.7/0/0/1,-0.2/0/0/0}{
      \node[entry] at (2.12,\y) {$\a$};
      \node[entry] at (3.40,\y) {$\b$};
      \node[entry] at (4.68,\y) {$\c$};
    }

    \draw[line width=0.75pt] (1.56,-0.65) rectangle (5.30,2.93);

    \node[annot,align=center] at (6.12,1.0)
      {extended\\threshold};
    \node[font=\normalsize,inner sep=0pt] at (3.6,-2.04)
      {$\cG=\cH^{x_0,0}|_{\mathcal{X}'}$};

\end{tikzpicture}
\end{minipage}

\caption{Schematic of the dichotomy induction.}
\label{fig:dichotomy}
\end{figure}

Put $\cG\coloneqq\cH^{x_0,1-b}|_{\cX'}$. Since $x_0\in\cX'$, we have $\cG=(\cH|_{\cX'})^{x_0,1-b}$, so $\Ldim(\cG)\ge(k-1)\ell$, and $\cG$ is a submatrix of $\cH$. Apply the induction hypothesis to $\cG$. If \ref{it:a} holds for $\cG$, it holds for $\cH$. Otherwise $\cG$ contains an extended threshold of size $k-1$: points $z_1,\dots,z_{k-1}\in\cX'$ and hypotheses $f_0,\dots,f_{k-1}\in\cG$ with $f_i(z_j)=1$ iff $i<j$. Choose $g_i\in\cH^{x_0,1-b}$ with $g_i|_{\cX'}=f_i$. Then
\begin{enumerate}[label=(\roman*)]
\item $g_i(z_j)=1$ iff $i<j$;
\item $g_i(x_0)=1-b$ for all $i$;
\item $h(x_0)=h(z_j)=b$ for all $j$, since $x_0,z_j\in\cX'=h^{-1}(b)$.
\end{enumerate}
If $b=1$, then $x_0,z_1,\dots,z_{k-1}$ and $h,g_0,\dots,g_{k-1}$ form an extended threshold of size $k$ in $\cH$. If $b=0$, then $z_1,\dots,z_{k-1},x_0$ and $g_0,\dots,g_{k-1},h$ form an extended threshold of size $k$ in $\cH$. In either case \ref{it:b} holds.
\end{proof}

\paragraph{Deriving a quadratic-exponent bound.}
Let us briefly discuss how \cref{lem:dich} implies the bound $\Ldim(\cH)\le 2^{O(\Ldim^*(\cH)^2)}$. We prove this by showing via an induction on $t$ that for every $t\ge 0$ and nonempty class $\cH\subseteq \{0,1\}^\cX$, if $\Ldim(\cH)\ge2^{\binom{t+1}{2}}$, then $\Ldim^*(\cH)\ge t$.

For $t=0$, the binomial $\binom{t+1}{2}$ evaluates to zero, so $\Ldim(\cH)\ge1$ and $\Ldim^*(\cH)\ge0$. Let $t\ge1$ and suppose $\Ldim(\cH)\ge2^{\binom{t+1}{2}}=2^t\cdot2^{\binom{t}{2}}$. Apply \cref{lem:dich} with $k\coloneqq2^t$ and $\ell\coloneqq2^{\binom{t}{2}}$.

In case \ref{it:a}, there are a submatrix $\cG$ of $\cH$ and $g\in\cG$ with $\Ldim\bigl(\cG|_{g^{-1}(b)}\bigr)\ge2^{\binom{t}{2}}$ for both $b\in\{0,1\}$. By the induction hypothesis, $\Ldim^*\bigl(\cG|_{g^{-1}(b)}\bigr)\ge t-1$ for both values of $b$, so $\Ldim^*(\cG)\ge t$ by \eqref{eq:dualrec}, and therefore $\Ldim^*(\cH)\ge\Ldim^*(\cG)\ge t$ by monotonicity.

In case \ref{it:b}, $\cH$ contains an extended threshold of size $2^t$, and so $\Ldim^*(\cH)\ge t$ by \cref{fact:thr} applied to $\cH^*$.

\begin{remark}
The above proof also shows that if $\Ldim(\cH)\ge k\cdot2^{\binom{t}{2}}$ and $\Ldim^*(\cH)<t$, then $\tdim(\cH)\ge k$, since case \ref{it:a} of \cref{lem:dich} is excluded.
\end{remark}

\bibliographystyle{alpha}
\bibliography{bibs}
\end{document}